\documentclass[lettersize,journal]{IEEEtran}
\usepackage{amsmath,amsfonts}
\usepackage{algorithmic}
\usepackage{algorithm}
\usepackage{array}
\usepackage{textcomp}
\usepackage{stfloats}
\usepackage{url}
\usepackage{verbatim}
\usepackage{graphicx}
\usepackage{cite}
\usepackage{array}  
\usepackage{graphicx}  
\usepackage{makecell}
\usepackage[shortlabels]{enumitem}
\usepackage{caption}
\usepackage{subcaption}
\usepackage{float}

\usepackage{soul}
 \usepackage{array}
 \usepackage[longtable]{multirow}
 \usepackage{longtable}
 \usepackage{multirow}
 \usepackage{tablefootnote}
\usepackage{makecell}
\usepackage[flushleft]{threeparttable}
\usepackage{soul, color}

\begin{document}

\title{Toward quantum-safe scalable networks: an open, standards-aware key management framework}

\author{
\IEEEauthorblockN{
    Ane Sanz$^{1,2}$,
    Asier Atutxa$^{1,2}$,
    David Franco$^{1}$,
    Jasone Astorga$^{1,2}$,
    Eduardo Jacob$^{1,2}$ and
    Diego López$^{3}$ 
  }%

ane.sanz@ehu.eus, asier.atutxa@ehu.eus, david.franco@ehu.eus, jasone.astorga@ehu.eus, eduardo.jacob@ehu.eus, diego.r.lopez@telefonica.com

\IEEEauthorblockA{$^{1}$\textit{Department of Communications Engineering, University of the Basque Country (UPV/EHU). Faculty of Engineering of Bilbao, Plaza Ingeniero Torres Quevedo, n.1, Bilbao, 48013, Spain.}}%

\IEEEauthorblockA{$^{2}$ \textit{EHU Quantum Center, University of the Basque Country (UPV/EHU). Faculty of Science and Technology, Barrio de Sarriena s/n, Leioa, 48940, Spain.}}%

\IEEEauthorblockA{$^{3}$ \textit{Telefonica Innovación Digital, Distrito Telefónica, Ronda de la Comunicación, S/N, 28050 Madrid,Spain}}%

\thanks{This is the accepted manuscript of an article published in IEEE Network (2025). https://ieeexplore.ieee.org/document/11299406. DOI: 10.1109/MNET.2025.3636584. Published by IEEE. Licensed under CC BY-NC-ND 4.0.}

}

\markboth{Journal of \LaTeX\ Class Files,~Vol.~14, No.~8, June~2025}%
{Shell \MakeLowercase{\textit{et al.}}: A Sample Article Using IEEEtran.cls for IEEE Journals}




\maketitle

\begin{abstract}
With the advent of quantum computing, the increasing threats to security poses a great challenge to communication networks. Recent innovations in this field resulted in promising technologies such as Quantum Key Distribution (QKD), which enables the generation of unconditionally secure keys, establishing secure communications between remote nodes. Additionally, QKD networks enable the interconnection of multi-node architectures, extending the point-to-point nature of QKD. However, due to the limitations of the current state of technology, the scalability of QKD networks remains a challenge toward feasible implementations. When it comes to long-distance implementations, trusted relay nodes partially solve the distance issue through the forwarding of the distributed keys, allowing applications that do not have a direct QKD link to securely share key material. Even though the relay procedure itself has been extensively studied, the establishment of the relaying node path still lacks a solution. This paper proposes an innovative network architecture that solves the challenges of Key Management System (KMS) identification, relay path discovery, and scalability of QKD networks by integrating Software-Defined Networking (SDN) principles, and establishing high-level virtual KMSs (vKMS) in each node and creating a new entity called the Quantum Security Controller (QuSeC). The vKMS serves the end-user key requests, managing the multiple KMSs within the node and abstracting the user from discovering the correct KMS. Additionally, based on the high-level view of the network topology and status, the QuSeC serves the path discovery requests from vKMSs, computing the end-to-end (E2E) relay path and applying security policies. The paper also provides a security analysis of the proposal, identifying the security levels of the architecture and analyzing the core networking security properties, and a performance evaluation that confirms the feasibility and scalability of the proposal.
\end{abstract}

\begin{IEEEkeywords}
QKD, Trusted Relay, SDN, network security, quantum communications.
\end{IEEEkeywords}

\section{Introduction}

\IEEEPARstart{A}{s} quantum technologies evolve, Quantum Key Distribution (QKD) stands as a leading technology for securing communications with unconditional security \cite{scarani2009security}. Unlike classical cryptographic mechanisms that rely on computational assumptions, QKD is based on quantum mechanics to generate cryptographic keys with information-theoretic security.

While the typical application scenario of QKD is point-to-point by nature, QKD networks interconnect multiple nodes over broader areas, enabling key delivery beyond direct optical links. However, the scalability of such networks remains a challenge, especially as the number of users, links, loops, and relay hops increase. Addressing scalability requires architectural solutions for network control and infrastructure integration strategies that reduce deployment complexity \cite{li2025integration}.

\begin{figure}[htbp]
	\centerline{\includegraphics[width=0.8\linewidth]{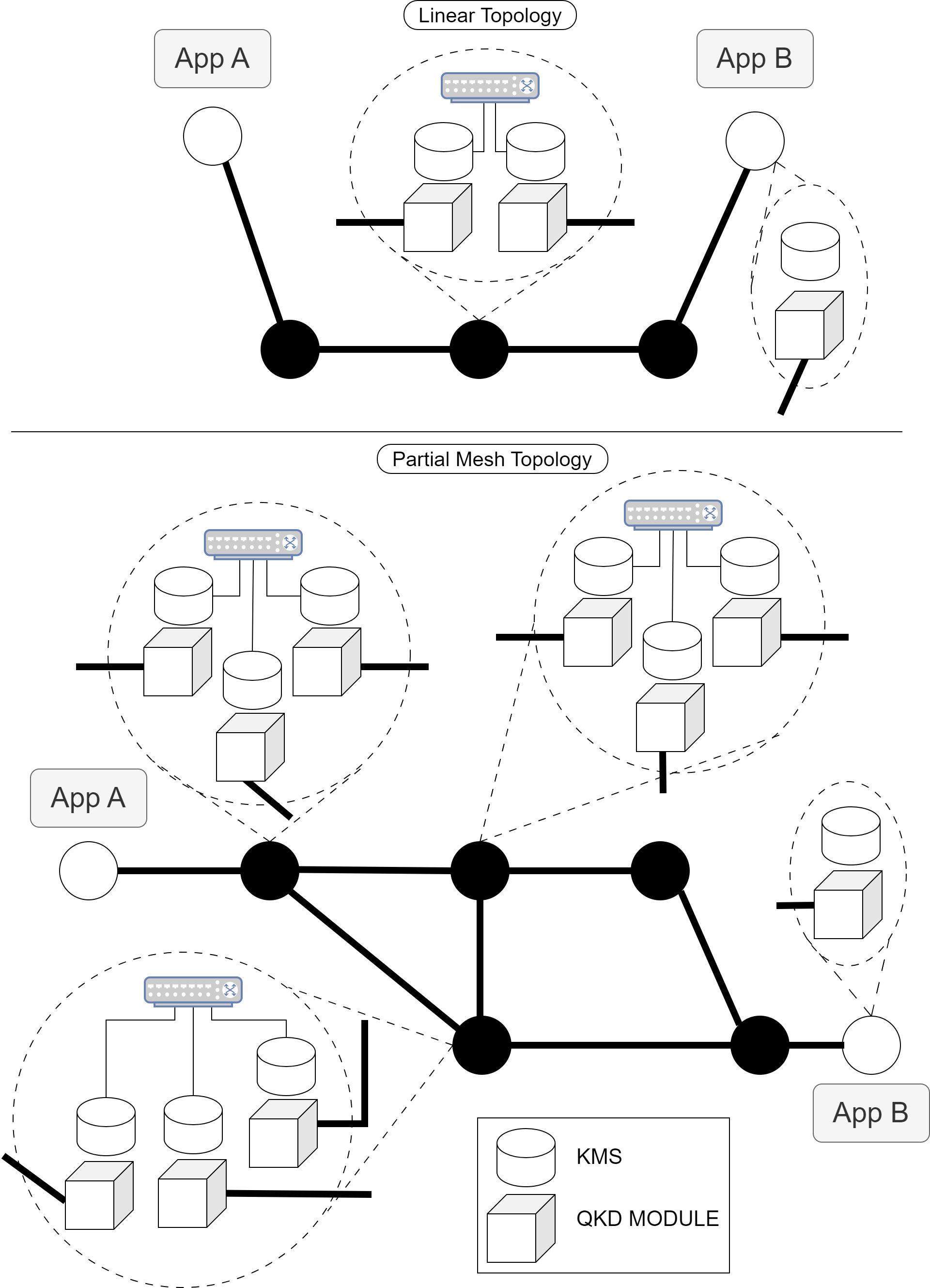}}
	\caption{A representation of linear (top) and partial mesh (down) topologies formed by QKD and trusted relay nodes.}
	\label{fig:qkd_network}
\end{figure}

QKD networks typically comprise two components. First, simple QKD nodes (white circles in Figure~\ref{fig:qkd_network}), each with a single QKD module and its associated Key Management System (KMS) responsible for managing, storing, synchronizing, and serving key retrievals. Second, trusted relay nodes (black circles in Figure~\ref{fig:qkd_network}), which host multiple QKD modules and KMS instances. These nodes serve as intermediaries, relaying keys across multiple hops to create virtual QKD links between non-directly connected endpoints.

Figure~\ref{fig:qkd_network} illustrates two QKD network topologies. The linear topology (top) represents most of the state-of-the-art trusted relay deployments, built pairs of Simple QKD nodes. This approach is applicable to static end-to-end (E2E) communications in which the relay path remains constant~\cite{ribezzo2023deploying} \cite{zhang2018large}. However, the partial mesh topology (down) illustrates a more realistic scenario, where multiple QKD nodes co-located in the same site enable the provision of dynamic relay paths.

In these networks, when an application initiates a secure session with a remote peer, its KMS must determine whether the peer is directly reachable or requires a multi-hop relay. This introduces several challenges in the design and operation of scalable QKD-based communications:

\begin{itemize}
    \item \textbf{KMS Identification:} In nodes hosting multiple KMSs, selecting the appropriate KMS to manage an E2E key request from applications requires additional coordination.
    \item \textbf{Path Discovery and Relay Coordination:} In nodes hosting multiple KMSs, Each KMS must identify and coordinate with the correct KMS instance within the same node to perform the relay of keys.
    \item \textbf{Scalability:}  The system must minimize the operational complexity exposed to applications and KMSs, while ensuring the architecture remains scalable as the number of nodes and key requests grows.
\end{itemize}

These issues stem from the limited knowledge at individual KMSs, and by the limitations of current QKD standards, which support basic functions but lack full dynamic multi-hop capabilities. To address these limitations, this paper proposes a secure multi-hop key delivery framework based on a centralized, programmable architecture leveraging Software-Defined Networking (SDN) principles. This approach abstracts complexity from applications and KMSs, introducing a global coordination mechanism to resolve paths and manage relay processes. The proposed design enables a centralized discovery and resolution of relay paths and remains compatible with existing and evolving QKD standards. Additionally, it enables seamless integration with different manufacturers and technologies.

The rest of the paper is organized as follows. Section \ref{sec:RelatedWork} summarizes the background and literature research. Section \ref{sec:ProposedSystem} describes the proposal Section \ref{sec:SecAnalysis} provides a security analysis and performance evaluation results of the proposed system. Section \ref{sec:conclusions} presents the main conclusions.

\section{Related work} \label{sec:RelatedWork}
A primary limitation of QKD is its restricted transmission distance typically 100–150 kilometers, due to signal attenuation and the lack of practical quantum repeaters. To overcome these constraints and support the deployment of scalable and complex network topologies, QKD networks commonly rely on trusted relay nodes, which have been widely studied \cite{cao2022evolution} \cite{tsai2021quantum}. 

Trusted relay nodes act as intermediate entities between distant endpoints, facilitating E2E key distribution by relaying keys across a chain of QKD links. At each hop, the key is decrypted and re-encrypted with a new key from the next link. While effective, this approach introduces challenges in routing, synchronization, and coordination, particularly in large or dynamic networks. While static routing approaches are used for smaller or stable networks, they lack the flexibility required in more dynamic environments.

Many QKD deployments are based on linear topologies, where each trusted relay node hosts two QKD modules and associated KMSs, making the relay procedure relatively straightforward. For instance, \cite{ribezzo2023deploying} presents a linear QKD network connecting cities in Italy, Slovenia, and Croatia through three links and two relays. Similarly, the Beijing-Shanghai Backbone QKD Network \cite{zhang2018large} connects 32 nodes linearly across 2.000 kilometers. While such networks demonstrate the feasibility of large-scale deployments, their linear nature limits scalability and flexibility.

These limitations become more pronounced in practical topologies where nodes may host multiple KMSs and QKD links. In such configurations, the process of discovering the best KMSs and computing an optimal relay path becomes significantly more complex. To address this, several efforts explore distributed routing approaches for QKD networks. For example, \cite{drif2025distributed} proposes the use of modified versions of classical routing protocols such as OSPFv2, where each node independently computes shortest path trees and maintains a local routing table to determine relay paths.

Other works pursue centralized architectures for QKD network control and key relay. For example, \cite{sasaki2011field} proposes a Key Management Agent per node and a centralized KMS coordinating all the links and networking functions.

In this context, SDN emerges as a promising approach. By decoupling control and data planes, SDN enables centralized control, programmability, dynamic path computation, and real-time reconfiguration, features well-suited for QKD networks with varying key rates, link availability, and security constraints. Several efforts validate SDN integration in QKD infrastructures. For instance, \cite{martin2024madqci} presents a network comprised of 28 QKD modules distributed across 9 nodes in Madrid. The network is coordinated by an SDN controller through the SDN Agents of the nodes. However, their relay nodes do not have KMS functionality for key retrieval, only extending the QKD reach. Recent studies \cite{lopez2025enhanced} have also explored virtualized control and key management models, enhancing flexibility and interoperability through softwarized architectures.

Additionally, \cite{bassi2023quantum} proposes an SDN-managed QKD infrastructure prototype where E2E keys are delivered via key relay procedure orchestrated by the controller, which computes suitable key relay paths. However, their description is based on a simple linear chain and the procedure of path computation is not described in detail, which limits its applicability to more complex topologies.

In line with these SDN-based approaches, the ETSI GS QKD 015 \cite{ETSI_GS_QKD_015} specification stands as a key enabler for building interoperable and programmable QKD networks. This specification defines an interface and data models between the Software-Defined QKD nodes and the controller, formalizing the architectural principles to integrate SDN control into QKD infrastructures.

To our knowledge, our proposal is the first to present a network architecture that logically centralizes the management of quantum security resources while performing the KMS discovery and end-to-end relay path in complex QKD network topologies.

\section{Proposed system} \label{sec:ProposedSystem}
This section introduces the proposed hierarchical network architecture along with its main components. It then describes the operational workflow of the proposed multi-hop key delivery mechanism. Finally, two representative use cases are presented to describe the operation and the detailed message exchange. 

\subsection{Architecture} 
The proposed architecture introduces a hierarchical KMS structure within QKD-enabled nodes to enhance scalability and simplify KMS discovery and relay operations. Each node incorporates a Virtual Key Management System (vKMS) that coordinates local key requests and a set of KMSs, each associated with a specific QKD link or module. This architecture integrates SDN principles through a new entity, the Quantum Security Controller (QuSeC), to enable centralized control and programmable behavior across the QKD network. The proposed architecture comprises the following key components, as depicted in Figure \ref{fig:steps}:

\begin{itemize}
    \item \textbf{Virtual KMS (vKMS):} It acts as the interface between all local applications and the QKD infrastructure. It abstracts the complexity of managing multiple QKD modules and local KMS instances, presenting a unified control point for key requests. The vKMS initiates the relay path discovery in coordination with the QuSeC.
    \item \textbf{Local KMS:} KMS instance associated to a physical QKD module with standard KMS functionalities: key storage, synchronization, management and exchange. It also handles both direct key retrieval and multi-hop key relaying in accordance with instructions received from the QuSeC. From this point forward, these instances will be referred to as simply KMS.
    \item \textbf{QuSeC:} It maintains a global view of the QKD network topology, including QKD links, node roles, and KMS associations. It responds to discovery requests from vKMS instances, computes optimal relay paths and enforces relay policies to all involved KMSs across the network. The relay path computation is performed based on SPF algorithm~\cite{dijkstra}, assigning the weights of each link according to configurable criteria such as available key rate, hop count, or physical distance, allowing for flexible adaptation to different optimization goals.
\end{itemize}


The architecture is agnostic to the underlying QKD protocol and transmission medium because all entities interact through standardized REST APIs (ETSI QKD 014 and 020), ensuring interoperability across heterogeneous vendors and devices. From a scalability perspective, the design supports horizontal scalability, enabling additional QKD nodes to be integrated without compromising the network and KMS operation. 

Beyond simple configuration of link weights, the QuSeC enables programmability through its northbound interface, allowing operators to program different routing or security policies dynamically. For instance, modifying the behavior of the Path Computation Element (PCE) in the QuSeC, the routing algorithm can be replaced or modified seamlessly.

An example of the proposed architecture is depicted in Figure \ref{fig:steps}. In this scenario, three nodes operate as trusted relay nodes, each hosting multiple KMS instances (each connected to a different QKD module, potentially from different vendors) to interface with separate QKD links. The fourth node acts as a simple QKD node with a single QKD module.

\subsection{Operational flow} 
\begin{figure*}[t]
\centering
\includegraphics[width=0.8\textwidth]{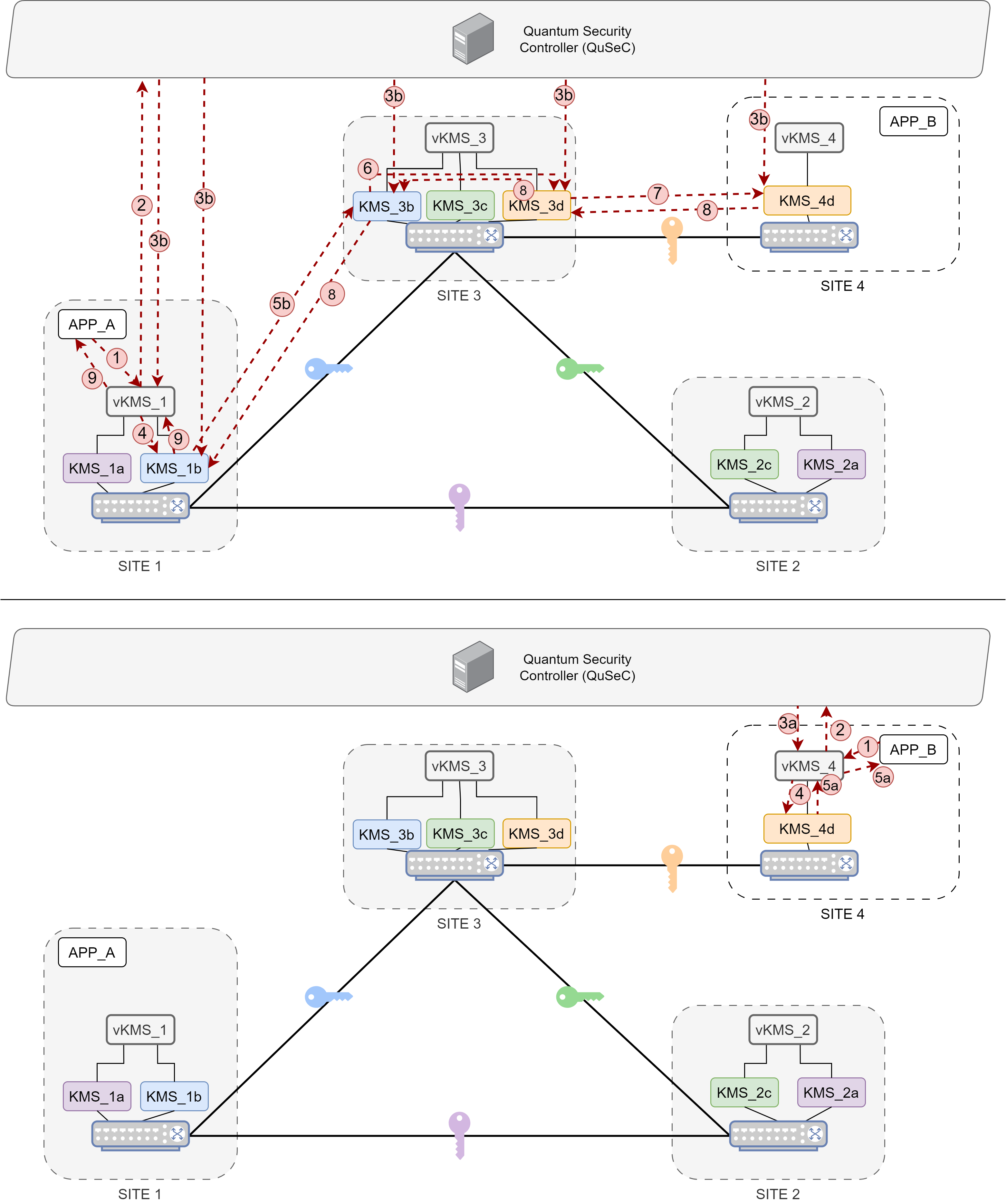}
\caption{Overall steps of the proposed mechanism for an initiating application (top) and target application (down).}
\label{fig:steps}
\end{figure*}

The proposed architecture enables E2E QKD key delivery through a mechanism that integrates centralized computation of the relay path. In addition, the mechanism are compliant with current specifications: ETSI GS QKD 014 \cite{ETSI_GS_QKD_014} for key delivery, ETSI GS QKD 020 \cite{ETSI_GS_QKD_020} for interoperable KMSs in horizontal key relay procedures, which is currently under development, and ETSI GS QKD 015 \cite{ETSI_GS_QKD_015} for SDN integration. This section outlines the main operational steps from the application's key request to the delivery of the final cryptographic key, as illustrated in Figure \ref{fig:steps}, which includes the operations required for initiating (top) and target (down) applications.

\begin{enumerate}
    \item \textbf{Application request to vKMS.} An initiating application (APP\_A) initiates a key request by sending a \texttt{Get Key} message to the vKMS of its node, using the ETSI GS QKD 014 API. The request includes the identifier of the target application (APP\_B). Applications only interact with their vKMS, which abstracts the complexity of the underlying QKD network.
    
    \item \textbf{KMS Discovery Phase.} Upon receiving the request, the vKMS sends a \texttt{KMS\_discovery\_request (APP\_A $|$ APP\_B)} to the QuSeC. This message includes the identities of both initiating and target applications. 
    
    \item \textbf{Path Computation and Installation by the QuSeC.} The QuSeC, based on its complete view of the network topology, computes the full relay path required to deliver the key from the initiating to the target application. Two cases arise:
    \begin{enumerate}
        \item  Direct path: if both applications are in the same QKD link domain or the requesting APP is the target app, the QuSeC identifies a single KMS responsible for delivering the key and responds with  a \texttt{KMS\_discovery\_response (APP\_A $|$ APP\_B $|$ ID\_KMS)}. In the latter case, since the QuSeC maintains the state of the computed relay path, no additional relay or path installation is needed for the target application (APP\_B).
        \item Trusted relay path: if intermediate hops are required, the QuSeC computes a relay path, determining a full ordered list of involved KMSs. For each KMS in the relay path, it sends a \texttt{relay\_path\_install(ID\_association $|$ previous\_hop $|$ next\_hop)}, allowing each KMS to locally configure its role in the relay chain, as defined in the ETSI GS QKD 015 specification. Then, the QuSeC returns a \texttt{KMS\_discovery\_response (APP\_A $|$ APP\_B $|$ ID\_KMS)} to the vKMS, identifying the first KMS in the relay path.
    \end{enumerate}
    

    \item \textbf{Key Retrieval Initiation.} The vKMS processes the QuSeC response and forwards the \texttt{Get Key} request to the corresponding KMS.
    
    \item \textbf{KMS Processing.} The KMS, upon receiving the \texttt{Get Key} request, performs the following:
    \begin{enumerate}
        \item Direct path: The KMS retrieves a QKD key from its storage (K1) and sends it to the vKMS, which forwards it to the application.
        \item Trusted relay path: The KMS, according to the relay path installed by the QuSeC, retrieves a QKD key from its storage (K1) and sends a \texttt{relay\_process\_request (APP\_A $|$ APP\_B $|$ ID\_relay\_key)} to its peer KMS to request for a relay process of the retrieved key. 
    \end{enumerate}
    \textit{Note: From this point forward, the procedure for the direct path is complete. The subsequent steps are applicable only in cases where a trusted relay is necessary.}
    
    \item \textbf{Key Relay Initiation.} The KMS that receives a \texttt{relay\_process\_request}, based on the \texttt{ID\_relay\_key}, retrieves the corresponding key value (K1) from its database. Then, based on the next hop of the relay path installed by the QuSeC, it will initiate an ETSI 020 \texttt{ext\_key\_request} message with the next KMS in the path, which includes the relay key ID and value of the key to be relayed.
    
    \item \textbf{Key Relay Forwarding.} Each KMS involved in the relay path performs the following steps:
    \begin{itemize}
        \item \textbf{When an \texttt{ext\_key\_request} is received:} It retrieves a QKD key (K2) from its local database and encrypts the  relay key (K1) using K2: $K3 = K1 \oplus K2$. Then, it sends the encrypted key and the identifier of the encryption key to its peer KMS. For this, a \texttt{key\_relay(encrypted\_relay\_key $|$ ID\_key\_encryption)} message is constructed and transmitted.
        \item \textbf{When a \texttt{key\_relay} is received:} The KMS retrieves K2 using \texttt{ ID\_key\_encryption} and decrypts K3 to recover the relayed key K1. According to the pre-installed relay path rules, if more hops remain, the KMS re-encrypts K1 with a new key, and relays it to the next hop, sending an \texttt{ext\_key\_request} message to the next KMS in the path.
    \end{itemize}
    This loop continues until the KMS that receives the relayed key has no next\_hop configured in the relay path, meaning that the final KMS in the path is reached, which receives and decrypts the key, storing it locally for delivery.

    \item \textbf{Relay Completion Notification.}  Once the key relay process ends, different response messages are sent back through the relay path:
    \begin{itemize}
        \item A \texttt{key\_relay\_response} is sent back from the KMS instances that received the relayed key to their peer KMS, determining the status of the relayed key reception and decryption.
        \item An \texttt{ack request 020} is sent from the KMS instances that relayed the encrypted key to the KMS instance that requested the relay. This message is defined in the ETSI GS QKD 020 specification draft.
        \item A \texttt{relay\_process\_response} is sent from the KMS that initiated the key relay to its peer KMS, from which it received the \texttt{relay\_process\_request}.
    \end{itemize}
   
    \item \textbf{Key Delivery to Application.} Once the initiating KMS receives the final key relay response, it forwards the key to the vKMS. The vKMS then delivers the key to APP\_A, completing the end-to-end key establishment.
    
\end{enumerate}




\begin{table*}[htbp]
\centering
\caption{Overview of messages and their purposes.}
\resizebox{\textwidth}{!}{%
\begin{threeparttable}
\begin{tabular}{|l|l|l|l|l|} 
\hline
\multicolumn{1}{|c|}{\textbf{Message}}    & \multicolumn{1}{c|}{\textbf{Parameters}} & \multicolumn{1}{c|}{\textbf{Sender $\rightarrow$ Receiver}} & \multicolumn{1}{c|}{\textbf{Standard}}  & \multicolumn{1}{c|}{\textbf{Purpose}}                                                                           \\ 
\hline
\texttt{Get Key} or \texttt{Get Key with ID}                & \texttt{APP\_dst}                                 & Application $\rightarrow$ vKMS                                & 014                                     & \makecell[l]{Requests a key for communication\\with destination application}                                                   \\ 
\hline
\multirow{2}{*}{\texttt{KMS\_discovery\_request}}  & \texttt{APP\_src}                                 & \multirow{2}{*}{vKMS $\rightarrow$ QuSeC}                      & \multirow{2}{*}{Proposed}               & \multirow{2}{*}{\makecell[l]{Requests KMS identifier for key\\establishment between applications}}                             \\
                                         & \texttt{APP\_dst}                                 &                                                               &                                          &                                                                                                                  \\ 
\hline
\multirow{3}{*}{\texttt{KMS\_discovery\_response}} & \texttt{APP\_src}                                 & \multirow{3}{*}{QuSeC $\rightarrow$ vKMS}                      & \multirow{3}{*}{Proposed}               & \multirow{3}{*}{\makecell[l]{Returns KMS identifier for key\\establishment between applications}}                              \\
                                         & \texttt{APP\_dst}                                 &                                                               &                                          &                                                                                                                  \\
                                         & \texttt{ID\_KMS}                                  &                                                               &                                          &                                                                                                                  \\ 
\hline
\multirow{3}{*}{\texttt{relay\_path\_install}}  & \texttt{ID\_association}                          & \multirow{3}{*}{QuSeC $\rightarrow$ KMS}                       & \multirow{3}{*}{Proposed}               & \multirow{3}{*}{\makecell[l]{Establishes relay path and\\configures all involved KMSs}}                                     \\
                                         & \texttt{prev\_hop}                                &                                                               &                                          &                                                                                                                  \\
                                         & \texttt{next\_hop}                                &                                                               &                                          &                                                                                                                  \\ 
\hline
\multirow{3}{*}{\texttt{relay\_process\_request}}  & \texttt{APP\_src}                                 & \multirow{3}{*}{KMS $\rightarrow$ Peer KMS}                   & \multirow{3}{*}{Proposed}               & \multirow{3}{*}{\makecell[l]{Initiates key retrieval for relay\\process}}                                                      \\
                                         & \texttt{APP\_dst}                                 &                                                               &                                          &                                                                                                                  \\
                                         & \texttt{ID\_relay\_key}                           &                                                               &                                          &                                                                                                                  \\ 
\hline
\multirow{5}{*}{\texttt{ext\_key\_request}}        & \texttt{ID\_relay\_key}                           & \multirow{5}{*}{KMS $\rightarrow$ KMS}                        & \multirow{5}{*}{020}                    & \multirow{5}{*}{\makecell[l]{Requests forwarding of key to\\next KMS in relay path}}                                        \\
                                         & \texttt{value\_relay\_key}                        &                                                               &                                          &                                                                                                                  \\
                                         & \texttt{APP\_src}                                 &                                                               &                                          &                                                                                                                  \\
                                         & \texttt{APP\_dst}                                 &                                                               &                                          &                                                                                                                  \\
                                         & \texttt{Others}\tnote{*}                                  &                                                               &                                          &                                                                                                                  \\ 
\hline
\multirow{2}{*}{\texttt{key\_relay}}               & \texttt{encrypted\_relay\_key}                    & \multirow{2}{*}{KMS $\rightarrow$ Peer KMS}                   & \multirow{2}{*}{Proposed}               & \multirow{2}{*}{\makecell[l]{Relays encrypted key to next\\KMS in path}}                                                       \\
                                         & \texttt{ID\_key\_encryption}                      &                                                               &                                          &                                                                                                                  \\ 
\hline
\texttt{key\_relay\_response}                      & \texttt{Status}                                   & KMS $\rightarrow$ Peer KMS                                    & Proposed                                 & \makecell[l]{Acknowledges status of key relay\\process}                                                                        \\ 
\hline
\multirow{5}{*}{\texttt{ack\_request}}             & \texttt{ID\_relay\_key}                           & \multirow{5}{*}{KMS $\rightarrow$ KMS}                        & \multirow{5}{*}{020}                    & \multirow{5}{*}{\makecell[l]{Acknowledges successful key receipt\\and decryption}}                                             \\
                                         & \texttt{ack\_status}                              &                                                               &                                          &                                                                                                                  \\
                                         & \texttt{APP\_src}                                 &                                                               &                                          &                                                                                                                  \\
                                         & \texttt{APP\_dst}                                 &                                                               &                                          &                                                                                                                  \\
                                         & \texttt{Others}\tnote{*}                                  &                                                               &                                          &                                                                                                                  \\ 
\hline
\texttt{relay\_process\_response}                  & \texttt{Status}                                   & KMS $\rightarrow$ Peer KMS                                    & Proposed                                 & \makecell[l]{Final status response to relay\\process request}                                                                  \\
\hline
\end{tabular}
\begin{tablenotes}
    \item[*] \textit{Only relevant parameters for this proposal are represented. All the parameters defined in the ETSI GS QKD 020 specification should be included.}
    \end{tablenotes}
\end{threeparttable}
} 
\label{tab:messages}
\end{table*}

Table~\ref{tab:messages} summarizes the main protocol messages exchanged between entities during the direct and relay-based key establishment.

\subsection{Message Exchange/examples}
This section presents two representative message exchange sequences under different scenarios: one using a direct key delivery and the other involving a trusted relay-based relay path. These examples focus on the interaction between the applications, vKMS, KMSs, and the QuSeC during the key request and delivery process.

\begin{figure*}[t]
\centering
\includegraphics[width=1\textwidth]{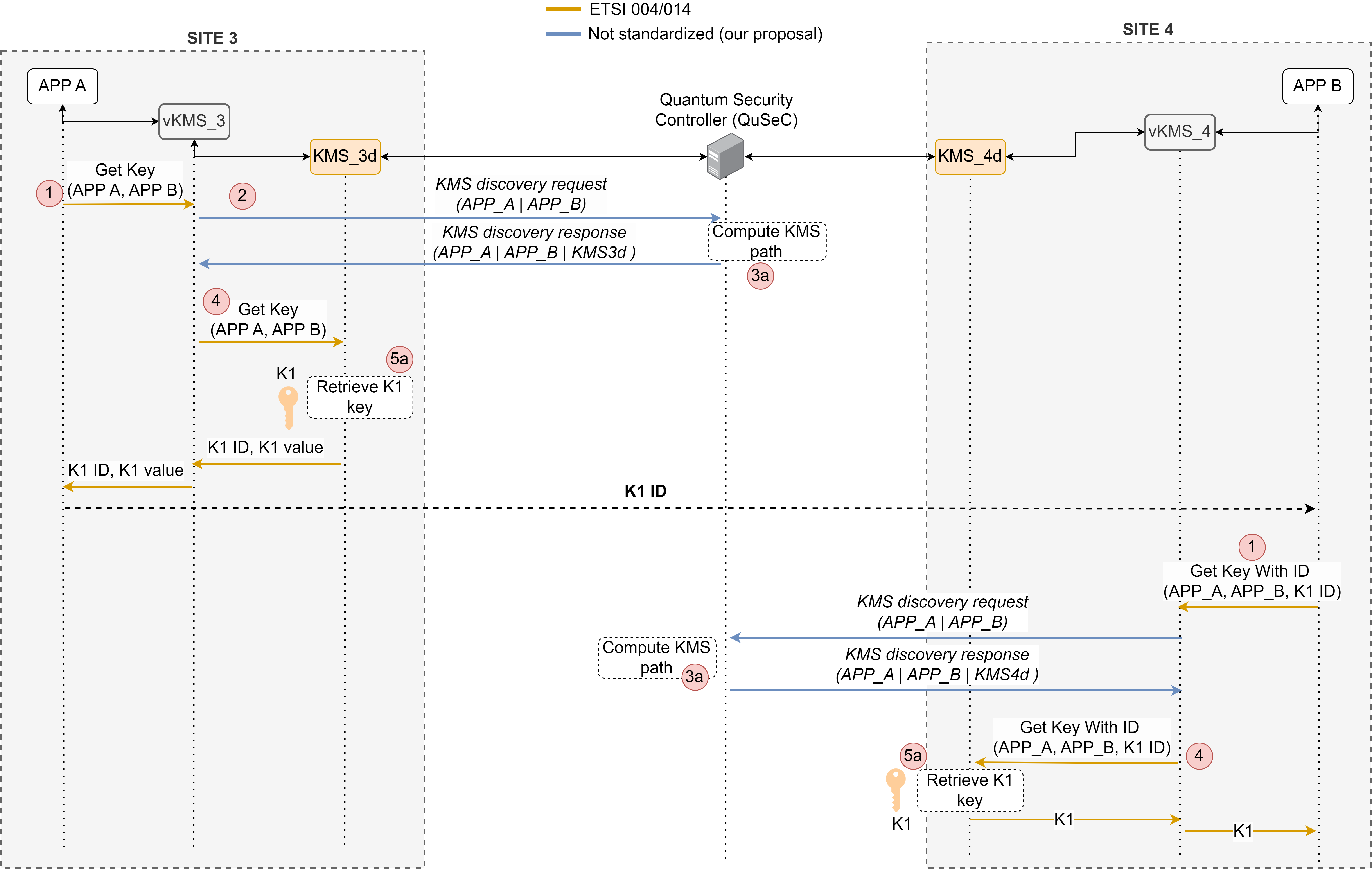}
\caption{Message exchange for direct path use case.}
\label{fig:direct}
\end{figure*}

On the one hand, Figure \ref{fig:direct} shows the message flow for a \textbf{direct path} use case, where both applications (APP\_A and APP\_B) reside within the same QKD link domain (KMS\_3d-KMS\_4d). In this case, for APP\_A, the QuSeC determines that a direct delivery is possible and responds with the KMS identifier, allowing the vKMS to forward the \texttt{Get Key} request directly to KMS\_3d. The key is retrieved and delivered without invoking any relay operation. For APP\_B, the process is symmetric: it sends a \texttt{Get Key with ID} request to vKMS\_4, which queries the QuSeC to discover the identity of KMS\_4d, forwards the request, and retrieves the matching key. The mechanism by which the key ID is shared between applications is outside the scope of this work.

\begin{figure*}[t]
\centering
\includegraphics[width=1\textwidth]{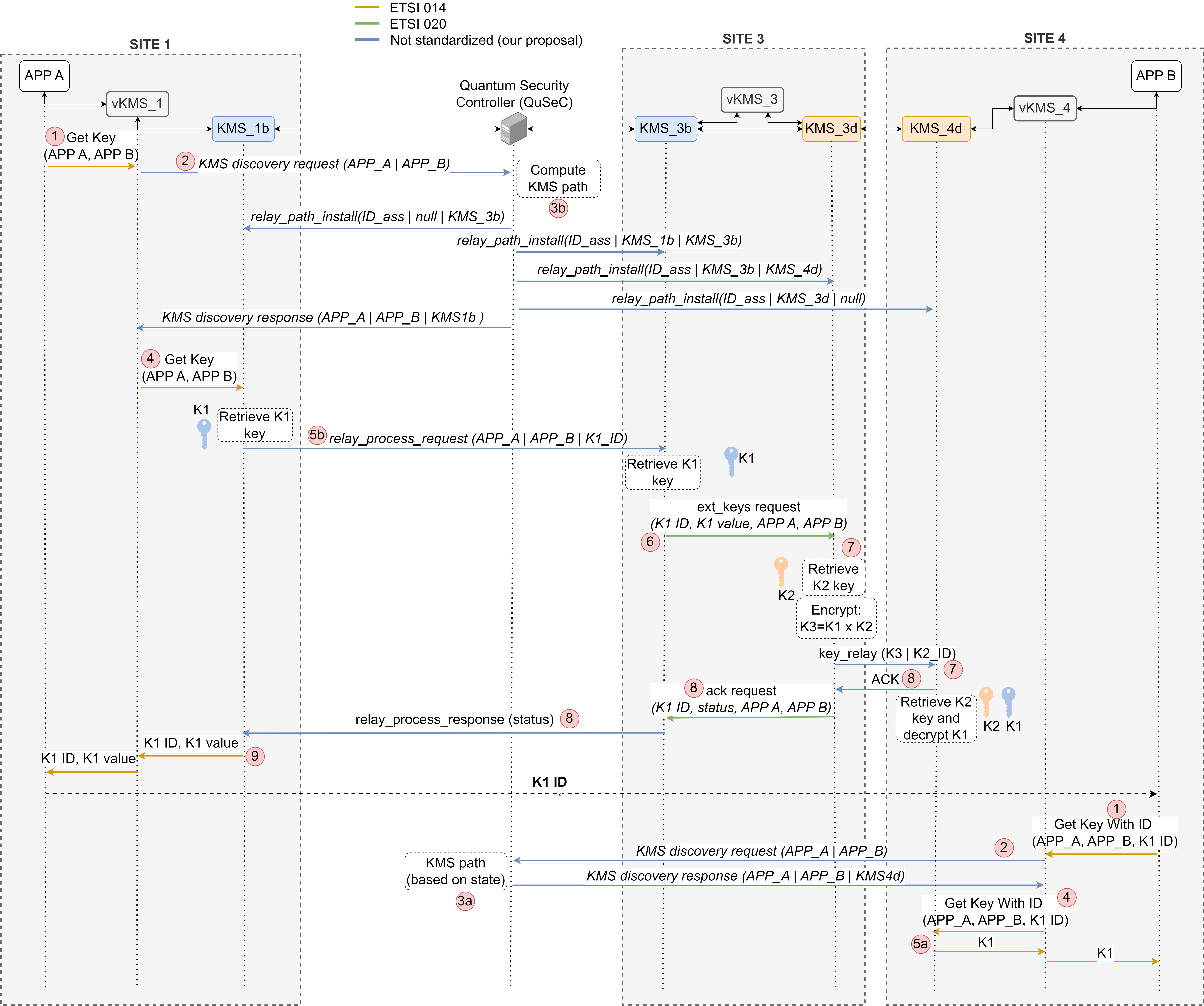}
\caption{Message exchange for single-hop trusted relay use case.}
\label{fig:1hop}
\end{figure*}

On the other hand, Figure \ref{fig:1hop} depicts the message exchange for a \textbf{single-hop trusted relay} case, where the initiating and target application are in different QKD link domains and a single trusted relay node is required. Upon receiving the discovery request from vKMS\_3, the QuSeC computes the full relay path, installs the necessary rules across involved KMSs \{KMS\_1b, KMS\_3b, KMS\_3d\, KMS\_4d\}, and returns the ID of the starting KMS to the vKMS. The vKMS then forwards the \texttt{Get Key} message to KMS\_1b, which initiates the relay process. 

KMS\_1b retrieves a QKD key (K1) and sends a \texttt{relay\_process\_request} to its peer, KMS\_3b. The latter retrieves K1, and following the installed relay path rules, sends an \texttt{ext\_key\_request} to KMS\_3d. This KMS retrieves another key (K2) and encrypts the incoming key with it ($K3 = K1 \oplus K2$). It then forwards the encrypted key and the ID of K2 to KMS\_4d in a \texttt{key\_relay} message. KMS\_4d retrieves K2 and decrypts K3 to recover K1 and stores it locally.

Since no further hops are installed in the KMS, the relay process is completed. In addition, completion notifications are propagated back through the chain. Once KMS\_1b receives the confirmation of successful relay completion, it responds to the initial \texttt{Get Key} request, returning K1 to the vKMS, which then delivers it to APP\_A. 

APP\_B’s behavior mirrors the direct case: it sends a \texttt{Get Key with ID} request to its local vKMS. Since the QuSeC has already installed the required path and stores state information, the vKMS receives from the QuSeC a simple KMS identifier and forwards the request directly to the KMS (KMS\_4d), which already holds the decrypted key.

\section{Security analysis and performance assessment}
This section presents a theoretical security analysis of the proposed framework and the results of the performance evaluation.

\subsection{Security analysis} 
\label{sec:SecAnalysis}
To perform the validation, we first present several assumptions that mark the fundamentals of the analysis. We adopt the Dolev-Yao adversary model, where the adversary is able to intercept, modify, and change any message on the classical channels, either those used by the QKD algorithms, those used for key relying, or those in the control plane, although it cannot break the cryptographic primitives. We also assume that the quantum channels are secure, and that the nodes that are part of the architecture are physically secure. Apart from that, we comply with the ETSI QKD specifications, and therefore the communications between entities, namely QuSeC, vKMS, KMS, and applications, are protected through mutual authenticated and encrypted channels.

\subsubsection{Architectural security}
The presented architecture proposes a hierarchical design of key management. Here, there are several security layers where components are classified: KMSs, vKMSs, and QuSeC, with different contributions and functions:

\begin{itemize}
    \item \textbf{Local KMS:} Each QKD link endpoint is equipped with a local KMS responsible for the key storage, reconciliation, and delivery. The communications between the KMSs and the QKD devices are verified at link level, as each KMS is bounded to its QKD device. The security features of the KMSs are the following:
    \begin{itemize}
        \item Link confidentiality: The quantum channels are secure and the transmitted bits are kept confidential, only sending them as part of relay procedures through encryption.
        \item Link authentication: The KMS pair communicates through a mutually authenticated channel, ensuring protection against man-in-the-middle.
        \item Isolation of keys per link: In sites with more than one link, and therefore multiple QKD devices, each device has an isolated KMS, not sharing communication channel and storage. 
    \end{itemize}
    \item \textbf{Virtual KMS (vKMS):} It is an abstraction of the multiple KMSs, located in an upper logical level. The vKMS manages the application queries to the QKDs, in domains with more than one KMS, performing the required KMS discovery procedure. This decision is aligned with the ETSI QKD 020 specification. Security features of vKMS: 
    \begin{itemize}
        \item Unified authentication point: The vKMS will not forward keys or respond to requests unless properly authenticated and authorized, preventing attackers from gaining illegitimate access to cryptographic material.
        \item Policy enforcement and monitoring: The vKMS may enforce security policies within the domain toward the applications. These policies include limitations to request rates or key usage policies. 
        \item Interoperability across vendors: The vKMS provides an abstraction from the specific implementation of the KMSs and the manufacturers of the QKD devices toward the applications. With this layer, the application requests and communication are manufacturer-independent, reducing the risk of bugs and errors, supporting application mobility and avoiding vendor lock-in. 
    \end{itemize}
    \item \textbf{QuSeC:} The highest level of security and management hierarchy, analogous to an SDN controller for the QKD network. It is compliant with the ETSI QKD 015, and it is responsible for orchestration, relay path selection, and relay policy definition. Security features of QuSeC:
    \begin{itemize}
        \item End-to-end key orchestration: It determines the relay path between two applications that do not share a direct QKD link, establishing the route and installing the required relay instructions on each KMS. The path establishment follows specific security policies defined in the QuSeC by the administrator, using authenticated channels with the KMSs. This centralized management prevents any adversary from routing the keys, unless the QuSeC itself is compromised (which is assumed unfeasible).
        \item Authenticated control plane: The communication between the QuSeC and the KMSs/vKMSs is secured, following the standard APIs and corresponding authentication and confidentiality measures. Thus, any attacker’s messages that try to alter the policies or security routes of a KMS/vKMS are detected, as the cryptographic systems used for the authentication and encryption of the communication are deemed secure. 
        \item No exposure of key material: The QuSeC does not retrieve, see or manipulate any QKD key material, as the applications directly access the vKMS to request these secrets. Even though the QuSeC is deemed secure, if it would be compromised, the separation of control and key manipulation prevents the attacker from obtaining the keys. 
        \item Global policy and compliance enforcement: The QuSeC may enforce global security policies, applied to the relay path establishments or to the KMS/vKMS in each site. These policies may include avoiding specific combinations of QKD pairs (in large architectures with many sites), or even more advanced strategies such as path redundancy.
    \end{itemize}
\end{itemize}

\subsubsection{Security properties}
This section analyzes the core security properties of the proposed architecture and the protection against possible attacks.

\begin{itemize}
    \item \textbf{Confidentiality and key secrecy:} The objective of an adversary may be to learn the secret keys being distributed or relayed. However, the proposed system is correctly protected against this vulnerability. First, the keys are generated over a quantum channel, which by definition is information-theoretically secure. Thus, an attacker cannot eavesdrop the keys from the quantum channel without being detected, which causes the protocol to abort and the obtained key to be discarded. Furthermore, during the key relay process, the forwarded key is encrypted throughout the secure path, following the ETSI QKD 020 and 014 specifications using HTTPS with TLS and mutual authentication, and One-Time-Pad (OTP) encryption of the relayed key with a fresh QKD key is used. Additionally, the KMS systems are considered secure and trusted storage entities, which is in line with the typical assumptions regarding QKD technology.
    \item \textbf{Authentication and integrity:} The objective of an adversary may be to impersonate a legitimate entity or modify messages to deceive the system. In this context, The proposed architecture is designed to authenticate and protect the integrity of each critical message. This is achieved by performing mutual authentication between the entities, which is specified in the QKD specifications. In fact, an application that leverages the ETSI QKD 014 API to access a vKMS previously performs a TLS handshake, as well as between KMSs that leverage the ETSI QKD 020 API to perform a key relay. Thus, the authentication is robust as long as the used cryptographic systems are robust, such as certificates based on post-quantum cryptography. The integrity of the messages is also automatically provided after establishing a secure session using TLS, using sequence numbers and HMAC.  Additionally, key relay is performed through OTP, which inherently provides authentication and integrity, as the encryption is one-use-only and the source is authenticated. 
\end{itemize}

\subsection{Performance assessment}
\begin{figure}[t]
	\centering
	\includegraphics[width=1\columnwidth]{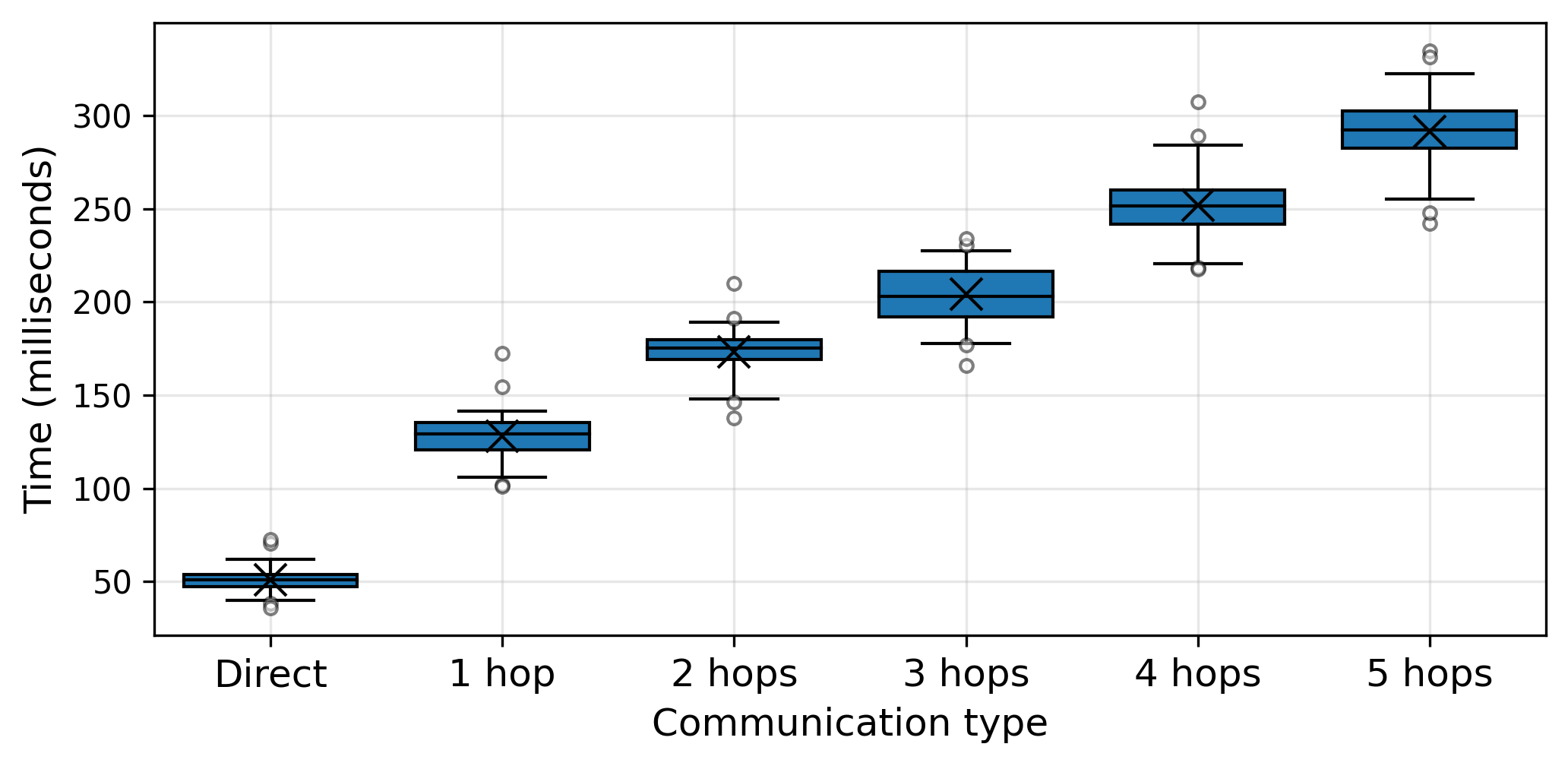}
	\caption{Measured E2E key establishment delay.}
	\label{fig:delay}
\end{figure}

To validate the feasibility and performance of the proposed architecture, a testbed was deployed on a Kubernetes cluster hosting containerized instances of the QuSeC, vKMS, and KMS entities. Figure \ref{fig:delay} presents the measured E2E key establishment delay for different communication types, ranging from direct path to five-hop relay scenarios. The path computation assumes equal link costs, so the routing criterion corresponds to the minimum hop count, although these weights could be dynamically adjusted to reflect metrics such as available key rate.

The results, averaged over 100 iterations per case, show a quasi-linear increase in delay with the number of hops, from about 50 ms for direct paths to around 290 ms for five hops. The most significant increase occurs between the direct and one-hop cases, as the first relay introduces both the QuSeC path-installation phase and the relay procedure itself, whereas additional hops only contribute with incremental encryption/decryption and messaging operations. These results confirm that the centralized control logic introduces negligible overhead compared with direct key exchange and that the relay mechanism scales efficiently even under multi-hop configurations.

\section{Discussion and Conclusions} \label{sec:conclusions}
This work has presented a complete architecture and mechanism for E2E quantum key establishment across QKD networks, combining centralized path computation with decentralized key relay execution. By abstracting the key management complexity through a virtual KMS layer and introducing an SDN-based quantum security controller, the proposal enables seamless key delivery over both direct and multi-hop trusted relay paths.

The design ensures interoperability and compliance with ETSI specifications, building on the 014 API for key delivery, 015 for SDN-based control, and anticipating integration with the emerging 020 for KMS-to-KMS key relay. The centralized QuSeC efficiently computes relay paths based on configurable metrics such as hop count or available key rate and maintains session state to avoid redundant configurations.

While this proposal follows a logically centralized model, distributed control approaches could enhance resilience by avoiding a single point of failure. However, QuSeC centralization provides global visibility and simplified interoperability with existing standards. This approach also remains scalable under high-load conditions with frequent key requests, since control-plane traffic for KMS discovery and path setup is orders of magnitude lower than data traffic. In large-scale deployments, QuSeC replication and state synchronization across redundant instances (using east/westbound interfaces defined for centralized controllers) could provide load balancing and high availability, mitigating performance bottlenecks and avoiding single-point failures.

Regarding performance, experimental results from demonstrate the feasibility of the proposal, showing E2E key establishment time scales efficiently with the number of hops. Overall, the results demonstrate that centralized control introduces negligible overhead and that caching at the vKMS level could further reduce latency and controller load for repeated path computations.

Overall, the proposed framework provides a practical, standards-aligned solution for E2E key delivery across QKD infrastructures. It supports a broad range of topologies, simplifies application integration, and remains extensible for future developments in KMS interoperability and controller intelligence.

\section*{Acknowledgments}
This work was supported in part by the European Comission through the project GN5-2 HORIZON-INFRA-2024-GEANT-01-SGA in the \textit{Working Group 6 Task 1 - Technology}, and in part by the Spanish Ministry of Science and Innovation in the project EnablIng Native-AI Secure deterministic 6G networks for hyPer-connected envIRonmEnts (6G-INSPIRE) (PID2022-137329OB-C44).

\bibliographystyle{IEEEtran}
\bibliography{references}

\newpage

\section{Biography Section}
 


\vspace{11pt}

\begin{IEEEbiographynophoto}{Ane Sanz}
received her BSc degree in Telecommunication Engineering in 2019 and her MSc degree in Telecommunication Engineering in 2021 from the University of the Basque Country (UPV/EHU). She joined the Communications Engineering Department of the UPV/ EHU as a researcher in the I2T Research Lab (Engineering and Research on Telematics) in 2018. Her research interests include quantum communications applied to security and networking. Currently, she is a PhD student in Mobile Network Information and Communication Technologies at the UPV/EHU.
\end{IEEEbiographynophoto}

\begin{IEEEbiographynophoto}{Asier Atutxa}
    received his BSc and MSc degrees in Telecommunication Engineering in 2018 and 2020, and his PhD in Mobile Network Information and Communication Technologies from the University of the Basque Country (UPV/EHU) in 2025. He joined the Communications Engineering Department of the UPV/EHU as a researcher in the I2T Research Lab (Engineering and Research on Telematics) in 2018, where he currently works. His research interests include cybersecurity, software-defined networks, and virtualization applied to industrial environments. 
\end{IEEEbiographynophoto}

\begin{IEEEbiographynophoto}{David Franco}
    received the B.Sc. and M.Sc. degrees in Telecommunication Engineering from the University of the Basque Country (UPV/EHU) in 2016 and 2018, respectively. He also received the Ph.D. in Mobile Network Information and Communication Technologies from the UPV/EHU in 2025. He joined the UPV/EHU in 2016 as a researcher in the I2T (Engineering and Research on Telematics) research group, and he has been a PhD Assistant Professor in the Communications Engineering Department of UPV/EHU since 2025. His research focuses on software-defined networking and network function virtualisation applied to traffic engineering and secure communication systems.
\end{IEEEbiographynophoto}

\begin{IEEEbiographynophoto}{Jasone Astorga}
    received the B.Sc. and M.Sc. degrees in Telecommunication Engineering and the Ph.D. degree from the University of the Basque Country (UPV/EHU), in 2004 and 2013, respectively. From 2004 to 2007, she worked with Nextel S.A., a telecommunications enterprise. She joined the UPV/EHU in 2007 as a lecturer and a researcher with the I2T Research Laboratory. She is currently an Assistant Professor with the UPV/EHU. She has led and contributed to numerous local, national, and European research projects in cybersecurity for industrial environments, Software Defined Networking (SDN), Network Function Virtualization (NFV), and IoT security. She has co-authored numerous journal papers and conference contributions, and has supervised four PhD candidates and numerous master's and bachelor projects.
\end{IEEEbiographynophoto}

\begin{IEEEbiographynophoto}{Eduardo Jacob}
\{Senior Member, IEEE\} received the M.S. in Industrial Engineering  and Ph.D. degree in Information and Communication Technologies in 2001 from the University of the Basque Country (UPV/EHU), Spain. He is currently Full Professor with the Department of Communications Engineering at UPV/EHU. He is the I2T Research Laboratory principal investigator and SmartQuanT4E and SmartNets4E research infrastructures director. His research interests include computer networks, network security, and advanced network architectures, with a focus on quantum safe and time-sensitive networking. He has authored and co-authored numerous publications in international journals and conferences and actively participates and lead national and European research projects.\end{IEEEbiographynophoto}

\begin{IEEEbiographynophoto}{Diego Lopez}
received the M.S. in Physics from the University of Granadada, and the Ph.D. in Physics from the University of Seville in 2000. He joined Telefonica in 2011 as a Senior Technology Expert, where he is currently in charge of the Technology Exploration activities within the GCTIO Unit. Before joining Telefónica he spent some years in the academic sector, dedicated to research on network services, and was appointed member of the High-Level Expert Group on Scientific Data Infrastructures by the European Commission. Diego is currently focused on applied research in network infrastructures, with a special emphasis on data-driven management, new architectures, security, and quantum communications. Diego is an ETSI Fellow and chairs the ETSI TC DATA and ISG ZSM.\end{IEEEbiographynophoto}


 \end{document}